\documentclass[a4paper,fleqn,usenatbib]{mnras}
\usepackage{newtxtext,newtxmath}
\usepackage[T1]{fontenc}
\usepackage{ae,aecompl}

\title[Statistical mechanics of galaxy]{Statistical mechanics of gravitating gas like galaxy}

\author[A. Kashuba]{Alexander B. Kashuba,$^{1}$\thanks{E-mail: alexander$_-$kashuba@yahoo.com}}

\date{Accepted XXX. Received YYY; in original form ZZZ}

\pubyear{2017}

\begin{document}
\label{firstpage}
\pagerange{\pageref{firstpage}--\pageref{lastpage}}
\maketitle

\begin{abstract}

The most probable state of an infinite self-gravitating gas in the dynamical equilibrium is defined by `gravitational haziness', a parameter representing many-body effects and formally like the temperature in the case of thermal equilibrium. A kinetic equation is constructed using a concept of statistical equipartition of the virial among subsystems of the self-gravitating gas. A closed equation for the gravitational potential is conjectured as a special property of the kinetic equation. An equilibrium particle distribution function in the phase space, an analog of the Maxwell-Boltzmann weight, and a galaxy equation of state are found for all `gravitational haziness'. The first law of a `hazydynamics' (thermodynamics) states that the total mass of an astronomical stellar collection is the sum of the Archimedes displaced mass and an excess `gobbled' mass determined by the `gravitational haziness' and history.

\end{abstract}

\begin{keywords}
galaxies: kinematics and dynamics, elliptical and lenticular, cD

\end{keywords}

\section{Introduction}

Stellar dynamics, governed by the Newtonian gravity, has developed over the last century into a matured subject \cite{Saslaw}. Noticeably, Jeans equation guides us through complicated anisotropic kinematic properties of star clusters as discovered by Oort \cite{Saslaw}. Zwicky has used the virial equilibrium to evaluate the dark matter in galaxies \cite{Zwicky}. Recently, computer simulations \cite{Hut} reveal novel and important details of the stellar dynamics like the binary star formation, the core oscillation, the mass segregation, the star evaporation and the core collapse \cite{Spitzer}. Yet, a microscopic statistical mechanics capable to predict macroscopic properties of the globular clusters, elliptical galaxies and clusters of galaxies is lacking. Instead, astronomers are still relying on the empirical de Vaucouleurs law for the galaxy surface brightness whereas wealth of data on globular clusters \cite{McMaster} is coded in terms of the King's model \cite{King}.

This paper seeks to connect a more rigorous microscopic statistical mechanics theory of the self-gravitating gas and the astronomy. The random matrix theory \cite{Wigner,Dyson} and the spin glasses \cite{BY} are major examples of strongly correlated systems with all round interactions irrespective of the distance. In the stellar dynamics a similar collective phenomenon, the violent relaxation \cite{LB}, has been observed in numerical simulations. The state of a system in the dynamical equilibrium is distinct from the state in the thermal equilibrium \cite{Chavanis}. Building on that, this paper finds a kinetic equation with the kernel encoding both the strong correlations between and the equipartition of the Clausius virial among subsystems of the self-gravitating gas. An equilibrium particle distribution function in the phase space, the solution of the kinetic equation and analog of the Maxwell-Boltzmann weight, is found to depend on a gravitational haziness, a variable formally like the temperature. Unlike common view as being finite, galaxies in this paper are infinite pressurized intrinsically and disposed for an expansion. The implications for the astronomy are far-reaching. For elliptical galaxies and globular clusters, a new equation of state, an analog of the Clapeyron gas equation, is found. The surface brightness and the anisotropic velocity profiles, new relations for rotation vs size, the core mass vs the overall mass and the surface brightness vs pressure as well as the super-massive black hole in the centre of a galaxy \cite{SBH,SBH2} are all discussed in this paper.

\section{Kinetic Equation}

The gravitational potential near a moving particle, a star, of the self-gravitating gas can be divided into a smooth part $\Phi(\vec{r})$ and a time dependent part $\phi^{(1)}(\vec{r}(t))$ rapidly fluctuating in the co-moving frame. Many particles contribute to both potentials. For simplicity, let the gas of the size $R$ consists of $N$ particles with the same unit mass. Among particles, the nearest neighbor, constantly changing its identity, applies the random force with a typical dispersion:
\begin{equation}
\langle \left( \vec{\nabla}\phi^{(1)}(t) \right)^2 \rangle \sim G^2 /a^{4} \sim \frac{1}{N} G \rho(\vec{r}) \Phi(\vec{r}) \frac{R}{a},
\label{randomNearest}
\end{equation}
where $a$ is the mean distance between the nearest particles. The average gravitational potential and the average density $\rho(\vec{r})$ are related by the Poisson's equation:
\begin{equation}
\vec{\nabla}^2\Phi(\vec{r})+4\pi G\rho(\vec{r})=0,
\label{Poisson}
\end{equation}
where $G$ is the Newton constant. In statistical mechanics I find convenient to invert the sign of the gravitational potential. All particles in the gas can be divided into $N$ spheres of influence with respect to a given particle. $k$-th sphere of influence comprises approximately $k$ particles at the distance $k^{1/3} a$ away. The amplitude of their random force depends on their mutual correlations while being isotropic on average. Inspired by unique cooperative phenomena when all parts of the system interact equally as described by the random matrix theory \cite{Wigner,Dyson} and the spin glass theory \cite{BY}, I assume that the random force incoming from each sphere of influence is approximately the same and, therefore, simply multiply Eq.(\ref{randomNearest}) by $N$. Also, the time it takes for a particle to cross the whole gas in a random wobbly trajectory is approximately the same as the period of a classical mechanics orbit in the smooth potential $\Phi(\vec{r})$ \cite{Schwarzschild}. Thus, the time-dependent fluctuating force has a mean square:
\begin{equation}
\langle {\nabla}^\alpha \phi^{(1)}(t) \nabla^\beta \phi^{(1)}(t') \rangle = \frac{4}{9} \sqrt{\Theta G  \rho(\vec{r})}\, \Phi(\vec{r}) \delta^{\alpha\beta} \delta(t-t')
\end{equation}
where $\Theta$ is a dimensionless, velocity diffusion constant controlling the meandering of the real trajectory away from the orbit. Throughout this paper it is called the gravitational haziness. In terms of the Langevin stochastic dissipative dynamics \cite{kadanoff} of a chosen particle:
\begin{equation}
\frac{d\vec{r}}{dt} = \vec{v}, \qquad \frac{d\vec{v}}{dt} = \vec{\nabla}\Phi(\vec{r}) + \vec{\nabla}\phi^{(1)}(t)+\vec{f}_{drag}(\vec{v},\vec{r}),
\label{Langevin}
\end{equation}
this fluctuation force is generated by the stochastic action \cite{MRS,DP}:
\begin{equation}
S[\phi^{(1)}]= \frac{9}{8\sqrt{\Theta}}\int dt \frac{1}{\sqrt{G\rho(\vec{r}(t))}\, \Phi(\vec{r}(t))}  \vec{\nabla}\phi^{(1)}\cdot \vec{\nabla}\phi^{(1)}.
\label{MRS}
\end{equation}
In neutral plasmas the inter-particle correlations are limited spatially to within the Debye radius and the Balescu Lenard kinetic equation determines the drag force. Similarly, for the self-gravitating gas the gravitational drag was found \cite{chandra}. Unlike the non-linear Balescu Lenard equation, the effective kinetic equation in the self-gravitating gas strongly correlated by all round interactions is intrinsically one-particle and linear. Remember that in spin glasses the effective Thouless Anderson Palmer equation describes one spin in the effective field and it solves the problem exactly \cite{BY}. Also, the equation for the density of states in the random matrix theory is linear too and is exact for the large sizes of matrices. 

In thermal equilibrium the gravitational drag and the random force fluctuations, Eq.(\ref{MRS}), are related by the fluctuation dissipation theorem \cite{FDT}, thus, completing the definition of the Langevin dynamics Eq.(\ref{Langevin}). However, in the dynamical equilibrium it is rather the virial equipartition that relates them. For the one-particle Langevin stochastic dissipative dynamics there are two effective fields: the momentum diffusion Eq.(\ref{MRS}) and the effective gravitational drag different from the Chandrasekhar's one. With all these in mind, the kinetic equation for the particle distribution function $\psi(\vec{r},\vec{v})$ in the phase space, the Fokker-Plank equation \cite{kadanoff}, reads: 
\begin{eqnarray}
0=\frac{\partial}{\partial t}\psi+\vec{v}\cdot \frac{\partial}{\partial \vec{r}}\psi+ \frac{d\Phi}{d\vec{r}} \cdot \frac{\partial}{\partial \vec{v}}\psi - \nonumber\\ \frac{2}{3} \sqrt{\Theta} \sqrt{G\rho(\vec{r}(t))} \frac{\partial}{\partial v^\alpha} \left( L^{\alpha\beta}(v,\Phi) \frac{\partial}{\partial v^\beta}\psi+ 2v^\alpha \psi\right),
\label{kinetic}
\end{eqnarray}
where the upper line is the Liouville's part and $L^{\alpha\beta}(v,\Phi)$ is the mean-field velocity diffusion kernel. Both $L$ and $\psi$ are unknown in Eq.(\ref{kinetic}) and this ambiguity will be dealt with by imposing physical conditions later on. The gravitational haziness $\Theta$ enters the kinetic equation in the same way as the temperature does for a gas composed of large heavy particles moving in the viscous gas of light particles and where the Stokes friction law applies. Perhaps similarly, stars are moving through a haze of the gravitational field fluctuations coupled to the density fluctuations. This form of the kinetic equation, Eq.(\ref{kinetic}), does allow for the direct exchange of the heat between the gas of particles and the viscous gas and the entropy of the gas of particles can change both way unlike the case of the Boltzmann H-theorem.

The divergence form of the last term in the kinetic equation, Eq.(\ref{kinetic}), ensures the conservation of the total number of particles, and hence the mass, of the gas:
\begin{equation}
\frac{\partial \rho}{\partial t} + \frac{\partial}{\partial \vec{r}}\cdot \rho\, \vec{u} =0.
\label{continuity}
\end{equation}
The particle distribution function is normalized such as to give the following average mass density and the average local velocity of the gas in motion:
\begin{equation}
\rho(\vec{r})= \int \psi(\vec{r},\vec{v})\ d^3\vec{v}, \qquad \rho(\vec{r})\, \vec{u}(\vec{r})= \int \vec{v} \psi(\vec{r},\vec{v})\ d^3\vec{v}.
\end{equation}
The most probable state of the self-gravitating gas in the dynamical equilibrium is the state in rest: $\vec{u}(\vec{r})=0$.

The kernel $L^{\alpha\beta}(v,\Phi)$, already representing effects of strong correlations between remote subsystems, can be defined using a concept of local virialization of gas subsystems. Unlike the Maxwell-Boltzmann gas, a single particle here is too small to form a subsystem. In the self-gravitating gas, a subsystem possesses a property of local density as the far away subsystems interact via the gravitational force. The Gibbs's statistical physics is founded on the concept of equipartition of the conserved properties, like energy, mass, charge, element, momentum or angular momentum, among all subsystems \cite{kadanoff}. Similarly, the total virial of the self-gravitating gas can be subdivided into the sum of virials of the local gas subsystems: 
\begin{equation}
2T+U=\int d^3\vec{r} \rho(\vec{r}) \left( \langle \vec{v}^2 \rangle (\vec{r}) - \frac{1}{2}\Phi(\vec{r})\right),
\label{virial0}
\end{equation}
where the double counting of the interaction is excluded. The mechanical global virial theorem states that $2T+U=0$. The equipartition of the virial is a stronger local statistical physics condition:
\begin{equation}
\langle \vec{v}^2 \rangle (\vec{r}) = \frac{1}{2}\Phi(\vec{r}).
\label{virial}
\end{equation}
For thick gravitational haziness, $\Theta\to\infty$, the last line in Eq.(\ref{kinetic}) should vanish separately. A simple kernel: $L^{\alpha\beta}=\delta^{\alpha\beta}\Phi(\vec{r})/3$, satisfies the virial theorem and the equipartition of the virial, Eq.(\ref{virial}). In this leading approximation, the particle distribution function is the Maxwell's ones: $\psi(\vec{v}) \sim \rho(\vec{r})\Phi^{-3/2}\exp(-3\vec{v}^2/\Phi)$. Erroneously, I searched for a solution of the kinetic equation with this kernel using Maxwell's PDF multiplied by polynomials in the velocity. Whole Eq.(\ref{kinetic}) does allow for such a solution order by order, up to the fourth order terms, $1/\Theta^2$, satisfying the gas in rest condition. However, the kernel $L^{\alpha\beta}(v,\Phi)$ has to be modified in the third order and such solution violates the time reversal symmetry and is nonphysical. In this exercise making wrong choices quickly leads to a contradiction whereas by imposing the following equation on the potential:
\begin{equation}
\vec{\nabla}^2 \Phi + \Phi^{\displaystyle 5+1/\Theta} +\frac{1}{4 \Theta} \frac{\left( \vec{\nabla} \Phi \right)^2}{\Phi}=0,
\label{GeneralDensity}
\end{equation}
the longest series can be constructed. It is simply the Poisson isotropic equation Eq.(\ref{Poisson}) for the gravitational potential where the density of the gas, $\rho(\vec{r})$, is the sum of the last two terms divided by $4\pi G$. By writing the density in this form the equation of state of the self-gravitating gas is stated implicitly. Despite Eq.(\ref{GeneralDensity}) depends on one parameter $\Theta$, the gravitational haziness, let the coefficient in front of the third term be independent. If it is zero then Eq.(\ref{GeneralDensity}) is the Lane-Emden equation with the exponent $k=5+1/\Theta$. For $1<k<3$, the Lane-Emden equation describes the thermodynamic equilibrium of a star of finite mass \cite{chandra}. For the relativistic equation of state at $k=3$, only the equilibrium at the Chandrasekhar mass limit is possible. Unlike that, the Lane-Emden equation for $5<k<\infty$ describes an infinite collection of mass with the potential being the sum of a regular and an oscillating parts. The regular part falls off at large distance as $r^{-2 \Theta/(1+4 \Theta)}$. The third term, if switched on, suppresses the oscillating part until it disappears altogether for Eq.(\ref{GeneralDensity}). At even stronger third term the solution describes a finite mass. In the parameter space, Eq.(\ref{GeneralDensity}) represents a special critical line. Its physical infinite isotropic solution is:
\begin{equation}
\Phi(\vec{r})=\left(1+\frac{1+4\Theta}{12\Theta} \vec{r}^2\right)^{\displaystyle -\frac{2\Theta}{1+4\Theta}},
\label{PhiProfile}
\end{equation}
for the average gravitational potential of the gas and
\begin{equation}
4\pi\rho(\vec{r})=\left(1+\frac{\vec{r}^2}{36\Theta}\right) \left(1+\frac{1+4\Theta}{12\Theta} \vec{r}^2\right)^{\displaystyle -2\frac{1+5\Theta}{1+4\Theta}},
\label{DensityProfile}
\end{equation}
for the average density of the gas. Both are functions of the re-scaled radius vector $\vec{r}$ and depend on the three parameters: the gravitational haziness $\Theta$, the core radius and the central density both set to one. For thick gravitational haziness, the density profile Eq.(\ref{DensityProfile}), shows also an intermediate crossover radius $r_{mid}=\sqrt{12\Theta}$, much larger than the core radius. Within $r_{mid}$ the density profile follows that of the Plummer model whereas the gravitational potential is the Newtonian one. The case of thin gravitational haziness, $\Theta\to 0$, though, is more close to the thermodynamic equilibrium. The dynamical equilibrium is intrinsically unstable and it evolves towards the thermal equilibrium eventually \cite{Chavanis}.

After this preliminary, a solution of the kinetic equation Eq.(\ref{kinetic}) can be written as the generating series: $\psi(\vec{r},\vec{v})=$
\begin{equation}
\int \frac{d^3\vec{q}}{(2\pi)^3} e^{i\vec{q}\cdot\vec{v}} \left( \rho -\frac{1}{2!}q_a q_bT^{ab}_{(2)} +\frac{i}{3!} q_a q_b q_cT^{abc}_{(3)} +... \right),
\label{series}
\end{equation}
where the coefficients depends on the coordinates inside the gas and represent the moments of the velocity distribution: 
\begin{equation}
T^{ab...c}_{(n)}(\vec{r})=\int v^a v^b ... v^c \psi(\vec{r},\vec{v})\ d^3\vec{v}.
\end{equation}
The first coefficient, $T^a_{(1)}(\vec{r})=\rho u^a$, is zero for the gas in rest. Eq.(\ref{kinetic}) with the kernel: $L^{ab}(v,\Phi)=\Phi(\vec{r})\, \delta^{ab}/3+$
\begin{equation}
+\frac{2}{21}\frac{\partial}{\partial v^c} \left( v^a v^b v^c v^d-\frac{\vec{v}^2}{6} A^{abcd}(\vec{v}) +\frac{\vec{v}^4}{12} B^{abcd} \right)\frac{\partial}{\partial v^d},
\end{equation}
annihilates the next odd coefficient: $T^{abc}_{(3)}(\vec{r})=0$. Here, the notations: $B^{abcd}=\delta^{ab}\delta^{cd}+ \delta^{ac}\delta^{bd}+ \delta^{ad}\delta^{bc}$ and
\begin{eqnarray}
A^{abcd}(\vec{x})=x^ax^b\delta^{cd}+ x^ax^c\delta^{bd}+ x^ax^d\delta^{bc}+ \nonumber\\ + \delta^{ab} x^cx^d+ \delta^{ac}x^bx^d+ \delta^{ad}x^bx^c,
\end{eqnarray}
where $\vec{x}$ is either $\vec{r}$ or $\vec{v}$, are used. Thus, the kinetic equation order by order of the velocity moments expansion is reduced to a sequence of simpler equations. The first of them, the hydrostatic equilibrium equation, reads:
\begin{equation}
\partial_a T_{(2)}^{ab}=\rho(\vec{r})\, \partial_b \Phi,
\label{hydrostatic}
\end{equation}
where the gradient of the Jeans pressure on the left hand side balances the gravitational pull on the right hand side. The solution of Eq.(\ref{hydrostatic}) satisfying the virial equipartition condition: $T_{(2)}^{aa}(\vec{r}) =\rho (\vec{r}) \Phi (\vec{r}) /2 $, reads:
\begin{equation}
T^{ab}_{(2)}(\vec{r}) =\frac{1}{6}\rho(\vec{r})\Phi(\vec{r})\frac{\displaystyle \delta^{ab}+r^a r^b/12\Theta}{\displaystyle 1+\vec{r}^2/36\Theta },
\label{anisotropy}
\end{equation}
where the gravitational potential and the density are given by Eqs.(\ref{PhiProfile},\ref{DensityProfile}). Divided by the density, the Jeans pressure tensor, Eq.(\ref{anisotropy}), gives the velocity dispersion as the function of coordinates. It reveals a kinematic anisotropy between the radial and the transverse velocity dispersions. While being equal inside $r_{mid}$ they drastically diverge for $r>r_{mid}$. On the outskirts of the gas particles follow the high apogee orbits. Therefore,  $r_{mid}$ may be called the kinematic anisotropy radius, $r_{anis}=r_{mid}$. For thin gravitational haziness, $\Theta\to 0$, the average particle motions degenerate into the straight radial ones almost everywhere. Projected velocity dispersion profiles for elliptical galaxies have been measured using the planetary nebula spectrograph \cite{R} in qualitative agreement with Eq.(\ref{anisotropy}).

The second reduced kinetic equation reads:
\begin{equation}
\partial_d T_{(4)}^{abcd}= T_{(2)}^{ab} \partial_c \Phi + T_{(2)}^{ca} \partial_b \Phi + T_{(2)}^{bc} \partial_a \Phi.
\label{quartic}
\end{equation}
Its solution is not unique and we need some physical condition. For large distances away from the centre, $r\gg 1$, the stellar motions degenerates into the straight radial lines with a negligible velocity diffusion. Thus, the stationary and collisionless kinetic equation, the upper line of Eq.(\ref{kinetic}), applies at $r\gg 1$. Its solution depends largely  on the particle energy near the edge singularity of a semi-circular law:
\begin{equation}
\psi(\vec{r},\vec{v})=\frac{4\Theta}{(1+4\Theta)^2}\frac{1}{\pi\vec{r}^2} \sqrt{2\Phi(\vec{r})-v_r^2}\ \delta^{(2)}(\vec{v}_\perp),
\label{asymptote}
\end{equation}
where the particle energy is negative, $v_r^2<2\Phi(\vec{r})$, for closed orbits. Eq.(\ref{asymptote}) can be conveniently rewritten in the same representation as that of the series Eq.(\ref{series}):
\begin{equation}
\psi(\vec{r},\vec{v})= \rho(\vec{r}) \int \frac{d^3\vec{q}}{(2\pi)^3} e^{i\vec{q}\cdot\vec{v}} \frac{J_1\left( \sqrt{2\Phi(\vec{r}) \left( \vec{q}\cdot \vec{r} \right)^2/\vec{r}^2} \right)}{\sqrt{\Phi(\vec{r}) \left( \vec{q}\cdot \vec{r} \right)^2/2 \vec{r}^2}},
\label{condition}
\end{equation}
where $r\gg 1$ and $J_1(x)$ is the Bessel function. For arbitrary distances, the expansion of Eq.(\ref{condition}) in powers of $\vec{q}$ provides the physical condition to select a proper solution of Eq.(\ref{quartic}):
\begin{eqnarray}
T_{(4)}^{abcd}(\vec{r})=\frac{1}{6} \left(1+\frac{1+4\Theta}{12\Theta} \vec{r}^2\right)^{-3-\frac{6\Theta}{1+4\Theta}} \left( \frac{A^{abcd}(\vec{r})}{36\Theta} \right. \nonumber\\ \left. + \frac{1+4\Theta}{(12\Theta)^2} r^ar^br^cr^d + \frac{1+3\Theta +\Theta r^2}{3 (1+7\Theta)} B^{abcd}\right).
\label{fourthorder}
\end{eqnarray}
The particle distribution function in the dynamical equilibrium satisfies separately the Liouville's and the collision parts of the kinetic equation, like it does in the thermal equilibrium. Leaving the collision kernel $L$, that annihilates the odd velocity terms, unresolved, the particle distribution function in the dynamical equilibrium, an analog of the Maxwell-Boltzmann weight, can be found in all orders, using the Liouville's part alone, to consists of two terms:
\begin{eqnarray}
\psi(\vec{r},\vec{v})=\frac{1}{6\pi^2 (1+4\Theta)} \frac{1}{\sqrt{2\Phi(\vec{r})-\vec{v}^2- \frac{1+4\Theta}{12\Theta} [\vec{r}\times\vec{v}]^2}} + \nonumber\\ + \frac{1}{3\pi\sqrt{2\pi}} \frac{1+6\Theta}{1+4\Theta}  \frac{\Gamma\left(6+\frac{1}{\Theta}\right)}{\Gamma\left(\frac{9}{2}+\frac{1}{\Theta}\right)}  \left(\Phi(\vec{r})- \frac{\vec{v}^2}{2}\right)^{\displaystyle \frac{7}{2}+ \frac{1}{\Theta}},
\label{PDF}
\end{eqnarray}
where the gravitational potential is given by Eq.(\ref{PhiProfile}). Both terms depend only on the integrals of particle motion, the energy and the angular momentum, in accordance with the Jeans theorem. The first term vanishes for thick gravitational haziness whereas the last term vanishes for thin gravitational haziness. In the momentum space the particle distribution function develops an integrable divergence near a surface of an elongated ellipsoid marking the escape velocity.

Physically, this critical surface in the phase space seeds strong inter-particles correlations in the self-gravitating gas as was anticipated in the beginning of this section. Indeed, if there were no two-particle correlations then the usual Boltzmann kinetic collision integral would apply:
\begin{equation}
\textrm{St}\psi=AL\frac{G^2m\rho}{\sqrt{2\Phi}}\frac{\partial^2}{\partial \vec{v}^2} \psi=\frac{3}{4}AL\frac{G^2m\rho}{\sqrt{2\Phi}}\frac{1}{(\sqrt{2\Phi}-|\vec{v}|)^2} \psi,
\end{equation}
where $m$ is the typical mass of particles, $\rho$ is the central mass density, $A\sim 10$ is the constant, $L$ is the so-called Coulomb logarithm and $\vec{r}$ is in the centre for simplicity. We see that the rate of relaxation is quadratically divergent upon approaching the critical surface. This divergence overcomes any dilution of the gas and signals a growth of inter-particle correlations at least near the critical surface.

The hydrodynamic equation for the self-gravitating gas in motion, an analog of the Navier-Stokes equation:
\begin{equation}
\frac{\partial u^\alpha}{\partial t} +\vec{u}\cdot\vec{\nabla} u^\alpha +\frac{1}{\rho}\nabla^\beta t_{(2)}^{\alpha\beta} = \nabla^\alpha \Phi - \frac{4}{3} \sqrt{\Theta}\sqrt{G\rho} u^\alpha,
\label{NS}
\end{equation}
is dominated by the many-body gravitational drag, the last term, unlike the local one \cite{chandra}. Both the gravitational potential and the density depend on the motion in a non-linear way, see the Poisson's and the continuity equations Eqs.(\ref{Poisson},\ref{continuity}). The Jeans pressure: $T_{(2)}^{\alpha\beta}=t_{(2)}^{\alpha\beta}+\rho u^\alpha u^\beta$, though, is determined by a secondary equation:
\begin{eqnarray}
\frac{d t_{(2)}^{\alpha\beta}}{d t} + \partial_\gamma t_{(3)}^{\alpha\beta\gamma} + t_{(2)}^{\alpha\beta} \vec{\nabla}\cdot \vec{u} + t_{(2)}^{\alpha\gamma} \partial_\gamma u^\beta + t_{(2)}^{\beta\gamma} \partial_\gamma u^\alpha \nonumber\\ =\frac{8}{9} \sqrt{\Theta}\sqrt{G\rho} \left( \rho\left(3 u^\alpha u^\beta -\vec{u}^2 \delta^{\alpha\beta} \right) - t_{(2)}^{\gamma\gamma} +\frac{1}{2}\rho\Phi \right).
\label{NS1}
\end{eqnarray}
And so on. The gas motion induces the odd moments $T_{(2n+1)}$. For the self-gravitating gas the Navier-Stokes equation is actually a series of coupled equations encoding the equation of state in the dynamical equilibrium.

\section{Properties of gas in dynamical equilibrium}

In the remaining, the statistical mechanics of the self-gravitating gas will be projected into the night sky. Remember that Eqs.(\ref{PhiProfile},\ref{DensityProfile}) describe an infinite collection of mass in the dynamical equilibrium and it should merge into a uniform outer space density at large radius $R$. Galaxies are no different from industrial cylinders containing gas, they are either sealed or at ambient pressure. The stellar pressure is mediated by stars constantly joining or leaving a collection. For $\Theta\to\infty$, the Plummer model has a well defined mass confined in the core: $M_{core}=\sqrt{3}$. It is a fraction of the overall mass encircled by a large radius $R$:
\begin{equation}
M_{core}/M(R)=R^{\displaystyle -1/(1+4\Theta)}.
\label{centremass}
\end{equation}
In astronomy the situation when $M_{core}\approx M(R)$ is called the core collapse. In a typical galaxy, like large elliptical galaxy, the mass resides on the outskirts. For thin gravitational haziness, $\Theta\to 0$, the galaxy core becomes tiny. Alternatively, the total mass can be related to the outer space density $\rho(R)$ as:
\begin{equation}
M(R)=\frac{4\pi}{3}R^3\rho(R)\left(1+4\Theta\right),
\label{Glxmass}
\end{equation}
i.e. the Archimedes displaced mass times the gravitational haziness enhancement factor which, for $\Theta\to\infty$, may be capped at the logarithm of the ratio of $R$ to the core size. Mechanically, the total mass encircled by a large radius $R$:
\begin{equation}
G M(R)=\frac{4\Theta}{1+4\Theta}\, R\, \Phi(R),
\label{Gmass}
\end{equation}
is proportional to the gravitation potential on the outskirts $\Phi(R)$. The gravitational potential is a good measure of the line of sight velocity dispersion $\langle v_{los}^2\rangle(r) =\Phi(r)/6$, within the kinematic anisotropy radius $r_{anis}$, see Eq.(\ref{anisotropy}).

Let a stellar collection be confined by a spherical wall of radius $R$, much larger than the kinematic anisotropy radius $r_{anis}$. Let it exerts a pressure $p$ on the wall. In this case, the galaxy equation of state can be found from Eqs.(\ref{anisotropy},\ref{Gmass}):
\begin{equation}
M^2(R)=\frac{32\pi}{3 G}\ \Theta p R^4.
\label{galeq}
\end{equation}
It looks like the Clapeyron thermodynamic gas equation, $pV=NT$, although, the pressure here builds up for $\Theta\to 0$. The gravitational haziness is determined by the ratio of the total gravitational energy, which is not extensive in the thermodynamic sense, to the work needed to inflate the galaxy. Interestingly, at constant mass and gravitational haziness the galaxy equation of state follows the relativistic adiabatic law of the cosmic microwave background radiation.

Alternatively, a sealed stellar collection in vacuum can be sustained by a slow rotation of the gas as a whole rather than by ambient pressure. Inspecting the kinetic equation Eq.(\ref{kinetic}) there are two modifications in the rotating reference frame with a frequency $\Omega$. First, the gravitational force $\vec{\nabla}\Phi$ is augmented by the centrifugal and the Coriolis forces. Second, the virial is no longer $\Phi/2$ but rather $\Phi/2-\Omega^2\vec{r}^2$ and the force is no longer the gradient of the virial. Due to the prominence of the virial in the structure of the kinetic equation, the potential $\Phi(\vec{r})$ is better to be redefined into the virial. Then, the Poisson's equation reads:
\begin{equation}
\vec{\nabla}^2 \Phi + 4\pi G\rho\{\Phi,\Omega\}+8\Omega^2=0,
\label{GeneralOmega}
\end{equation}
where the gas density is now a functional of the virial $\Phi$ and the rotation frequency $\Omega$. The last term comes from the difference between the virial and the gravitational potential and it preserves the isotropy of Eqs.(\ref{Poisson},\ref{GeneralDensity}). The sign of this term is significant. If the series of the density functional in powers of $\Omega$ is weaker than this last term then Eq.(\ref{GeneralOmega}) describes a finite mass with a sharp boundary at some large outer radius $R(\Omega)$. For slow rotation, $\Omega\to 0$, it can be evaluated using the Eq.(\ref{GeneralDensity}) instead of the correct $\rho\{\Phi,\Omega\}$:
\begin{equation}
R(\Omega)\sim \Omega^{\displaystyle -(1+4\Theta)/(1+6\Theta)}.
\end{equation}
For thick gravitational haziness, $\Theta\to \infty$, the boundary outskirts of the self-gravitating gas rotate according to the Kepler's law whereas for thin gravitational haziness, $\Theta\to 0$, the boundary rotation velocity is constant. The elliptical galaxies do slowly rotate to keep stars with smaller and lighter ones rotating faster than the larger and heavier ones. The thicker is the gravitational haziness the slower large elliptical galaxies rotate. Unlike that and owing to living in a galaxy halo, globular clusters are sustained by an ambient halo density, and are typically spherical systems.

Let a compact self-gravitating gas be confined to within its core. For thick gravitational haziness this is possible, see Eq.(\ref{centremass}). Such an object is stable and needs neither ambient media no rotation. It has no outer radius and is completely defined by just two parameters: the central velocity dispersion, $\langle \vec{v}^2\rangle(0)$, and the central energy density, $\epsilon(0)$. To an outside observer it will look like a central mass, $M$, spreading out the Newton gravitation field $\Phi(r)=GM/r$. From Eqs.(\ref{PhiProfile},\ref{DensityProfile},\ref{anisotropy}) we find:
\begin{equation}\label{compact}
M=\sqrt{\frac{3}{\pi}}\frac{\langle \vec{v}^2\rangle^2}{G \sqrt{-G\epsilon}}, \qquad r_{core}=\frac{\langle \vec{v}^2\rangle}{\sqrt{-4\pi G\epsilon}}.
\end{equation}
Inspecting astronomical data on vastly disparate objects, from small cold molecular clouds near the Sun, through globular clusters orbiting in the Galaxy halo to huge elliptical galaxies, Eqs.(\ref{compact}) seem to be universal and the energy density $\epsilon$ is close to that of the cosmic microwave background radiation. It looks like a mechanical equilibrium. Almost as if the ambient cosmic microwave background radiation, a pump, pressurizes the stellar motions in a galaxy, a ball. Despite the Universe being apparently transparent the gravitational lensing phenomenon suggests that the cosmic microwave background radiation might apply a pressure and energize the Universe mechanically. Eq.(\ref{compact}), known as the Faber-Jackson law \cite{FJ},  relates the total luminosity of an elliptic galaxy to its line of sight velocity dispersion. For typical extended galaxy it is derived from Eqs.(\ref{Glxmass},\ref{Gmass}) with the energy density $\epsilon$ and the velocity dispersion, $\langle \vec{v}^2\rangle$, being defined on the outskirts:
\begin{equation}
M= \sqrt{\frac{3}{\pi}} \frac{(4\Theta)^{3/2}}{(1+4\Theta)^2} \frac{\langle \vec{v}^2\rangle^2}{G \sqrt{-G\epsilon}}.
\end{equation}
While for thick gravitational haziness the ambient pressure penetrates all the way towards the centre, a thin gravitational haziness system does resist mechanically. It balances the pressure on the boundary and may build up an internal pressure on its own.

The cores in nearby elliptical galaxies have been spatially resolved in observations and the surface brightness profile, proportional to the projected surface mass density, see Appendix, has been determined. Using Eqs.(\ref{galeq},\ref{surfbright}) the maximum central surface mass density:
\begin{equation}
\mu_{max}= \sqrt{\frac{32\Theta p}{3\pi G}} \left(\frac{1+4\Theta}{12\Theta} R^2\right)^{1-1/2(1+4\Theta)},
\label{mumax}
\end{equation}
is found to be proportional to the square root of the ambient pressure $p$. For thick gravitational haziness the central surface density as well as the central pressure, see Eq.(\ref{centerpressure}), increases more sharply with the galaxy size $R$ over the core size than that for the thin gravitational haziness.

A schematic and mechanistic picture of the galactic evolution follows from the statistical mechanics. It is possible that all galaxies were born non-rotating elliptical and sustained by the ambient pressure back then. As the Universe expands and the pressure drops the galaxies are expanding according to the equation of state, Eq.(\ref{galeq}), while some close by galaxies get mutual rotations. If remaining non-rotating, and hence blown up, galaxies are dispersing stars and feeding up the more compact rotating ones or, alternatively, settle down into the pressurized centre of a cluster of galaxies then the Darwin selection, survival of the rotating species, defines the galaxy evolution. If the pace of the Universe expansion is faster than a rotating galaxy can accommodate then it may puncture too. Dispersed stars, leaving the parent elliptical galaxy on the equator of the rotation, are burdened by the gravitation and form bending spiral arms that represent ordered exiting orbits in mechanical equilibrium as distinct from the remaining elliptical galaxy in the chaotic dynamical equilibrium.

This last example turns the statistical mechanics of self-gravitating gas into a `hazydynamics' confronted with questions like: if two close by galaxies have widely different gravitational haziness will they exchange something? does the gravitational haziness thickens towards the core? The laws of thermodynamics \cite{kadanoff} helps us in usual statistical physics. The first law of `hazydynamics' should relate masses by distinguishing them, unlike the Newton universal gravitation, in two categories (by rewriting Eq.(\ref{Glxmass})):
\begin{equation}
M_{total}=M_{Archimedes}+M_{excess}.
\end{equation}
The first category is either the Archimedes displaced mass or the mass sustained by the rotation or the outside pressure. Unlike that, the second category is an excess featureless gobbled mass, like the mass inside binaries. It is proportional to the gravitational haziness and keeps record of the history of the star collection. Also, dividing the mass into pieces will decrease the gravitational haziness whereas assembling the mass from dust lumps into stars in particular and the gravitational collapse in general will increase it.

At high-$z$ distance the pressure of the cosmic microwave background radiation decreases as the fourth power of the Universe scale. An elliptical galaxy in the dynamical equilibrium at constant mass and gravitational haziness expands coherently with the Universe, see Eq.(\ref{galeq}), and its surface brightness depends only on the gravitational haziness: $\mu\sim \sqrt{\Theta}$, not on $z$. For a disk galaxy, phase transitions and hydrodynamic instabilities may slow down this expansion. 

Recently, dramatic phenomena in many galaxies have been observed and interpreted as a super-massive black hole residing in the centre \cite{SBH2}. Nominal black hole is associated with the gravitational collapse in the end of the star evolution \cite{chandra}. The super-massive black hole is a different object and, as this paper may suggest, not necessary. It might be rather a stable self-gravitating gas of nominal black holes in dynamical equilibrium growing by the process of the mass segregation whereby heavy black holes sink into the centre. The reverse process of evaporation \cite{Spitzer} vanishes for $\Theta\to \infty$ and the cross section for a direct collision of two black holes is small. Such compact gas of black holes at infinite gravitational haziness is readily described by Eqs.(\ref{compact}). However, the energy density $\epsilon$ is by twenty or so orders of magnitude larger than that of the cosmic microwave background radiation. The super-massive black hole is found to be heavier in galaxies defined by thin gravitational haziness \cite{sbh}. The Kepler orbit of a probe star around the central object shows a perigee precession that depends on both $\Theta\neq \infty$ and $r_{core}\neq 0$ and is opposite to that of the Einstein general relativity effect.

\section{Conclusions}

The non-relativistic statistical mechanics theory of the self-gravitating gas in the dynamical equilibrium is developed. The particle distribution function in the phase space, an analog of the Maxwell-Boltzmann weight, and the associated equation for the gravitational potential are found using the proper kinetic equation. The first two hydrodynamic equations, an analog of the Navier-Stokes equation, are written down. All macroscopic static and kinematic properties of the self-gravitating gas depend on the gravitational haziness variable. The equation of state for a globular star collection in the dynamical equilibrium, like elliptical galaxy, is found. The physics of the core collapse phenomenon is clarified. For elliptical galaxies the relations between the rotation and the size, between the core and the overall masses and between the surface brightness and the pressure are all found. In the Appendix, the galaxy surface brightness profile, replacing the de Vaucouleurs law, is provided. A survey of the maximum surface brightness, Eq.(\ref{mumax}), and the gravitational haziness of non-rotating elliptical galaxies can possibly provide a map of the ambient pressure field in the local universe.

\appendix

\section{Surface brightness}

Close galaxies are clearly resolved with the internal structure being salient. In this case Eq.(\ref{DensityProfile}) is directly applicable. Distant galaxies, though, reveal light from the outskirt and can end in either sharp or a smooth manner. Lacking the quantitative theory of the rotating self-gravitating gas and the corresponding elliptical deformation I provide here two models for the surface brightness of a galaxy. First, the surface density, the mass projected along the line of sight, of a globular stellar collection given by the bulk density Eq.(\ref{DensityProfile}) with a sharp cutoff radius $R$ reads:
\begin{eqnarray}
\mu(r)=\frac{dM}{dA}(r)=\int^{\sqrt{R^2-r^2}}_{-\sqrt{R^2-r^2}}\rho\left(\sqrt{r^2+z^2}\right) dz = \nonumber\\ \frac{2}{3} \frac{\sqrt{R^2-r^2}}{\left(1+\frac{1+4\Theta}{12\Theta}r^2 \right)^{2+\frac{2\Theta}{1+4\Theta}}} \left[\left(\frac{12\Theta+(1+4\Theta) r^2}{12\Theta+(1+4\Theta) R^2} \right)^{\frac{1+6\Theta}{1+4\Theta}} + \right. \nonumber\\ \left. 2\left(1+\frac{r^2}{24\Theta}\right) F\left[\frac{1}{2},\frac{1+6\Theta}{1+4\Theta},\frac{3}{2},-\frac{(1+4\Theta)(R^2-r^2)}{12\Theta+(1+4\Theta)r^2} \right]\right],
\end{eqnarray}
where $r$ is the radius on the astronomical plate away from the centre and perpendicular to the line of site. $F[a,b,c,x]$ is the usual hypergeometric function. The second model adopts a soft density cutoff at large outer radius $R$:
\begin{eqnarray}
\mu(r)=\frac{dM}{dA}(r)=\int^{\infty}_{-\infty}\frac{\rho\left(\sqrt{r^2+z^2}\right)}{1+\left(r^2+z^2\right)/R^2} dz= \nonumber\\ \sqrt{\frac{12\pi\Theta}{1+4\Theta}} \frac{\Gamma\left(2-\frac{1}{2+8\Theta}\right)}{\Gamma\left(2+\frac{2\Theta}{1+4\Theta}\right)} \frac{\frac{R^2}{36\Theta}}{\left(1+\frac{1+4\Theta}{12\Theta}r^2 \right)^{2-\frac{1}{2+8\Theta}}} \times \nonumber\\ \left(1+\frac{36\Theta-R^2}{R^2+r^2} F\left[\frac{1}{2},1,-\frac{1+8\Theta}{2+8\Theta},\frac{12\Theta+(1+4\Theta)r^2}{(1+4\Theta)(R^2+r^2)} \right]\right) \nonumber\\ +\frac{\pi R^2}{\sqrt{R^2+r^2}} \frac{1-\frac{R^2}{36\Theta}}{\sin\frac{\pi}{2+8\Theta}}\frac{1}{\left(-1+\frac{1+4\Theta}{12\Theta}R^2 \right)^{2+\frac{2\Theta}{1+4\Theta}}},
\end{eqnarray}
where $R\gg r_{anis}$. For low-$z$ distances the surface brightness is the product of light to mass ratio, homogeneous in the dynamical equilibrium up to weak kinetic effects like the mass segregation, and the surface density. In both models, the surface density profile in the inner region, emerging in the limit $R\to\infty$, is the same:
\begin{equation}
\mu(r)=\mu_{max} \frac{1+r^2/24\Theta}{\left(1+(1+4\Theta)\, r^2/12\Theta \right)^{2-1/2(1+4\Theta)}},
\label{surfbright}
\end{equation}
where the maximum central surface density:
\begin{equation}
\mu_{max}=\sqrt{\frac{8\Theta p_{max}}{(1+4\Theta)G}} \frac{1+6\Theta}{1+4\Theta} \frac{\Gamma\left(1-\frac{1}{2+8\Theta}\right)}{\Gamma\left(2+\frac{2\Theta}{1+4\Theta}\right)},
\label{centerpressure}
\end{equation}
is determined by the central stellar Jeans pressure $p_{max}$. Remarkably, for the bulk density Eq.(\ref{DensityProfile}) this projected surface density is given by more or less the same formula. Moreover, there exists a potential in the two dimensions:
\begin{equation}
\Phi_{(2)}(\vec{r})=12\pi\Theta\mu_{max}\left(1+(1+4\Theta)\, \vec{r}^2/12\Theta \right)^{1/2(1+4\Theta)},
\end{equation}
such that this potential and the surface density are related by the two dimensional Poisson's equation:
\begin{equation}
\Delta^{(2)} \Phi_{(2)} -2\pi \mu(\vec{r})=0.
\end{equation}

\bsp
\label{lastpage}
\end{document}